# Quantum Entanglement of Free Particles

Roumen Tsekov

Department of Physical Chemistry, University of Sofia, 1164 Sofia, Bulgaria

The Schrödinger equation is solved for many free particles and their quantum entanglement is studied via correlation analysis. Converting the Schrödinger equation in the Madelung hydrodynamic-like form, the quantum mechanics is extended to open quantum systems by adding Ohmic friction forces. The dissipative evolution confirms the correlation decay over time, but a new integral of motion is discovered, being appropriate for storing everlasting quantum information.

According to quantum entanglement [1], when the initial quantum state of each particle cannot be described independently of the state of the others, the correlations persist at any time no matter if the particles are separated by large distances. The scientific literature is crowded by papers analyzing the entanglement of some internal properties of the particles [2, 3], e.g. spin or qubits, but in the present paper we are going to explore the quantum entanglement due to the momenta and positions of the particles. The analysis is based on the Schrödinger equation, which is well established in science. In contrast, the interpretations of Bohm and Nelson, for instance, are questionable and comparing their results with ours could elucidate some interpretation artefacts. Because the fundamental interactions vanish at large distance, the quantum entanglement must occur in pure form among non-interacting particles. This makes doubtful the traditional factorization of the wavefunction for weakly interacting particles [4], which is a key approximation in many quantum-mechanical methods such as HF, DFT, etc. When the particles are initially correlated, there is no way to factorize their wave function at any time, unless disturbing artificially the system to cause the wave function collapse, which is not the case in closed systems. There are firm indications, however, that in open quantum systems the friction force could destroy the entanglement due to entropy production. The latter is an information loss, which can be crucial for the sensitive modern quantum technologies by causing decoherence.

The joint wavefunction $\psi(\mathbf{x},t)$ of N free particles, each with mass $m$, obeys the following Schrödinger equation:

$$i\hbar\partial_t\psi = -\hbar^2\nabla^2\psi/2m \tag{1}$$

where $\nabla$ is the 3N-dimensional nabla operator. It is much easier to solve Eq. (1) via the Fourier transform, converting the Schrödinger equation $i\hbar\partial_t\phi = \mathbf{p}^2\phi/2m$ in the momentum representation. Here $\mathbf{p}$ is 3N-dimnsional vector of all momenta of all particles and $\mathbf{p}^2$ is the scalar product, which is equal to the sum of all momenta squares as expected in the kinetic energy of the entire system. For an initial Gaussian state, the solution of the momentum Schrödinger equation reads:

$$\phi(\mathbf{p},t) = \left|\pi\hbar^2\Sigma_0^{-1}/2\right|^{-1/4} \exp(-\mathbf{p}\cdot\Sigma_0\cdot\mathbf{p}/\hbar^2 - i\mathbf{p}^2 t/2m\hbar) \tag{2}$$

where $\Sigma_0$ is the initial covariance matrix of the positions $\mathbf{x}$ of all N particles. The momenta of the free particles are integrals of motion and the probability density $|\phi|^2 = f_0(\mathbf{p})$ is not changing in time. It means that the covariance matrix $<\mathbf{pp}> = \hbar^2\Sigma_0^{-1}/4$ of the momenta remains constant. Due to the unique properties of the Gaussian function, it is easy to obtain the covariance matrix of the positions at any time via $\Sigma^{-1} = [\Sigma_0 + (i\hbar t/2m)\mathbf{I}]_{\text{Re}}^{-1}$, where $\mathbf{I}$ is the 3Nx3N-dimensional unit tensor. The inversion of this expression leads straightforward to

$$\Sigma = \Sigma_0 + (\hbar t/2m)^2 \Sigma_0^{-1} \tag{3}$$

As a simple demonstration let us consider only two particles moving on a line. In this case the covariance matrix $\Sigma = \sigma^2 \begin{pmatrix} 1 & r \\ r & 1 \end{pmatrix}$ depends on the position dispersion $\sigma^2$ and the correlation coefficient $r$. Substituting this expression in Eq. (3) yields the evolution of the dispersion and correlation coefficient in time:

$$\sigma^2 = \sigma_0^2 + (\hbar t/2m)^2/\sigma_0^2(1-r_0^2) \qquad r = (2\sigma_0^2/\sigma^2 - 1)r_0 \tag{4}$$

If the two particles are initially not correlated ($r_0 \equiv 0$), they remain non-correlated at any time further, since $r = 0$. The spreading of the Gaussian wave packet in this case follows the well-known expression for a single particle. Any initial correlation between the two particles accelerates the spreading of the wave packet due to increase of the momentum dispersion. The latter is due to an additional initial potential, setting the correlation of the particles [5]. Since the correlation in the coordinate space corresponds to anticorrelation in the momentum space, the correlation coefficient will alter its sign at large time and $r_\infty = -r_0$. It means that the initial information will remain forever but there is an intermediate moment $\tau_1 \equiv 2m\sigma_0^2(1-r_0^2)^{1/2}/\hbar$, where the two particles are not correlated anymore. Hence, the probability density in the coordinate space factorizes at this moment. One can rewrite Eq. (4) in the dimensionless forms $\sigma^2/\sigma_0^2 = 1 + t^2/\tau_1^2$ and $r/r_0 = (1 - t^2/\tau_1^2)/(1 + t^2/\tau_1^2)$, respectively. Presence of everlasting spatial correlations among free particles can retain the initial shape of the body to maintain the fourth state of matter [6].

The description above refers to vacuum, where no energy dissipation occurs, and for this reason there is no entropy production as well. From practical point of view, it is essential to study the quantum entanglement under action of friction forces. Presenting the wavefunction in the Madelung polar form $\psi(\mathbf{x},t) = \rho^{1/2} \exp(iS/\hbar)$, where $\rho(\mathbf{x},t)$ is the probability density in the coordinate space, the Schrödinger equation (1) splits in two hydrodynamic-like equations:

$$\partial_t \rho = -\nabla \cdot (\rho \mathbf{v}) \qquad \partial_t \mathbf{v} + \mathbf{v} \cdot \nabla \mathbf{v} = -\nabla Q / m \tag{5}$$

The first is the compulsory continuity equation, where the hydrodynamic-like velocity $\mathbf{v} \equiv \nabla S/m$ represents the gradient of the wavefunction phase. The second equation is the momentum balance of the particles, where the quantum force is the gradient of the Bohm quantum potential $Q \equiv -\hbar^2 \nabla^2 \rho^{1/2}/2m\rho^{1/2}$ [7]. The latter is nonlocal, and it is an icon of the quantum entanglement because $-\nabla Q$ is the sole emergent force, acting among free particles. Hence, if there is any spooky action at a distance, it must occur through the Bohm quantum potential. When the particles are moving in an environment, the friction forces appear as well, which are proportional to the hydrodynamic-like velocity $\mathbf{v}$. Therefore, the force balance (5) changes correspondingly to

$$\partial_t \mathbf{v} + \mathbf{v} \cdot \nabla \mathbf{v} + \gamma \mathbf{v} = -\nabla Q / m \tag{6}$$

where $\gamma$ is the specific friction coefficient [8, 9]. The solution of Eq. (6), coupled to the continuity Eq. (5), is a normal distribution density $\rho = |2\pi\Sigma|^{-1/2} \exp(-\mathbf{x} \cdot \Sigma^{-1} \cdot \mathbf{x}/2)$, while the hydrodynamic-like velocity $\mathbf{v} = \dot{\Sigma} \cdot \Sigma^{-1} \cdot \mathbf{x}/2$ is linear. The correctness of these solutions can be proven by direct substitution in the equations above. Interestingly, the corresponding probability flow obeys the Fick law $\rho \mathbf{v} = -\dot{\Sigma} \cdot \nabla \rho / 2$ with a time-dependent diffusion tensor $\dot{\Sigma}/2$, which is typical for the Gaussian distribution. The evolution of the covariance matrix $\Sigma$ is governed by the following dynamic equation:

$$\ddot{\Sigma} - \dot{\Sigma} \cdot \Sigma^{-1} \cdot \dot{\Sigma}/2 + \gamma \dot{\Sigma} = (\hbar/m)^2 \Sigma^{-1}/2 \qquad (7)$$

The quantum force is linear as well, but $-\nabla Q = \hbar^2 \Sigma^{-2} \cdot \mathbf{x}/4m$ does not separate to independent contributions of each particle and even the force components of a single particle are correlated [10]. There are, however, some universal integrals of motion, e.g. $-<\Sigma \cdot \mathbf{x}\nabla Q/m> = (\hbar/2m)^2 \mathbf{I}$ [11]. In the case of non-correlated particles Eq. (7) reduces to $\ddot{\sigma} + \gamma \dot{\sigma} = (\hbar/2m)^2/\sigma^3$ and this dissipative Ermakov equation is already numerically solved [8].

In the case of vacuum ($\gamma \equiv 0$) the solution of Eq. (7) is the already derived expression from Eq. (3). In the opposite case of strong friction, one can neglect the first two inertial terms in Eq. (7) as compared to the friction force, and the solution of the remaining equation reads:

$$\Sigma^2 = \Sigma_0^2 + (\hbar^2 t/m^2 \gamma)\mathbf{I} \qquad (8)$$

Looking again for the two particles on a straight line with $\Sigma = \sigma^2 \begin{pmatrix} 1 & r \\ r & 1 \end{pmatrix}$, it follows from Eq. (8) that $r\sigma^4 = r_0 \sigma_0^4$. Therefore, the product of the correlation coefficient and the square of the dispersion is an integral of motion. This is very important because no matter of the strong friction the correlation $<x_1 x_2^3> = 3r\sigma^4$ is not changing in time. It could be used, for instance, to write quantum information for quantum computers, which does not decohere in a dissipative environment forever. Such conclusion is also valid for two different components of the position of a single particle. The evolution of the dispersion follows from Eq. (8) as well

$$\sigma^4(1+r^2) = \sigma_0^4(1+r_0^2) + \hbar^2 t / m^2 \gamma \tag{9}$$

Substituting here the relation $\sigma^4 / \sigma_0^4 = r_0 / r$ yields the evolution of the correlation coefficient:

$$r = \frac{1 + r_0^2 + t/\tau_2 - \sqrt{(1 + r_0^2 + t/\tau_2)^2 - 4r_0^2}}{2r_0} \tag{10}$$

where $\tau_2 \equiv \gamma(m\sigma_0^2 / \hbar)^2$ is the relaxation time, because $r$ tends to zero at $t \gg \tau_2$. Thus, the particles lose their initial correlation, which is expected because of the entropy production in the environment, but it happens very slowly at large friction. If the two particles are initially not strongly correlated ($r_0^2 \ll 1$) the relaxation time $\tau_2 = \gamma \tau_1^2 / 4$ corresponds to a harmonic oscillator with the own frequency $2/\tau_1$. In this case the correlation coefficient $r/r_0 = 1/(1 + t/\tau_2)$ and the dispersion $\sigma^2 / \sigma_0^2 = (1 + t/\tau_2)^{1/2}$ acquire simple dimensionless forms. At large time $t \gg \tau_2$ the product $rt = r_0 \tau_2$ remains constant, while the dispersion increases as the well-known expression $\sigma^2 = \hbar(t/\gamma)^{1/2} / m$ for a single particle [8]. The energy of the particle $\varepsilon = \hbar^2 / 2m\sigma^2 = \hbar\gamma / 2(\gamma t)^{1/2}$ decreases monotonously due to the friction in the surrounding. This expression has a clear meaning, while $\hbar\gamma / 2$ is the average quanta of energy exchanged during a collision of the quantum particle with particles from the environment, and $\gamma t$ is the number of collisions.

    The results obtained above are strictly valid for the Ohmic friction introduced in Eq. (6), which is an example of the Onsager thermodynamic relation between fluxes and forces, typical for slow processes. If the quantum particles are very fast, the friction force will depend nonlinearly on the hydrodynamic-like velocity $\mathbf{v}$ [12], and there will be probably other adiabatic integrals of motion, different form the product $r\sigma^4$ of the covariance and dispersion. Additional modulations could arise from thermal fluctuations in the surrounding. Neglecting the inertial terms in Eq. (6) yields the hydrodynamic-like velocity $\mathbf{v} = -\nabla Q / m\gamma$, which substituted in the continuity Eq. (5) leads to the nonlinear quantum diffusion equation $\partial_t \rho = \nabla \cdot (\rho \nabla Q) / m\gamma$ [8]. The latter is a particular example at zero temperature $T$ of the thermo-quantum diffusion equation:

$$\partial_t \rho = \nabla \cdot (\rho \nabla Q + k_B T \nabla \rho) / m\gamma \tag{11}$$

The effect of thermal fluctuations is to diminish the initial correlations, which is evident from the covariance matrix $\Sigma = \Sigma_0 + 2Dt\, \mathbf{I}$ for the classical diffusion, where $D \equiv k_B T / m\gamma$ is the Einstein diffusion constant. For two Brownian particles it follows $r/r_0 = 1/(1+t/\tau_3) = \sigma_0^2/\sigma^2$ with the classical relaxation time $\tau_3 \equiv \gamma m \sigma_0^2 / 2k_B T$.